\documentstyle[sprocl,epsfig]{article}

\def\be{\begin{equation}}
\def\ee{\end{equation}}
\def\bea{\begin{eqnarray}}
\def\eea{\end{eqnarray}}
\newcommand{\ccaption}[2]{
\begin{center}
\parbox{0.85\textwidth}{
\caption[#1]{\small {#2}}}
\end{center}    }

\def    \be             {\begin{equation}}
\def    \ee             {\end{equation}}
\def    \ba             {\begin{eqnarray}}
\def    \ea             {\end{eqnarray}}
\def    \nn             {\nonumber}
\def    \=              {\;=\;}

\def    \ie             {{\em i.e.,} }
\def    \eg             {{\em e.g.,} }

\def    \bra#1          {\mbox{$\langle #1 |$}}
\def    \ket#1          {\mbox{$| #1 \rangle$}}
\def    \dl#1           {$$\displaylines{\quad#1}$$}

\def    \({\left(}
\def    \){\right)}

\def    \mev            {\mbox{$\mathrm{MeV}$}}
\def    \gev            {\mbox{$\mathrm{GeV}$}}
\def    \pt             {\mbox{$p_T$}}

\def    \cf     {{\ifmmode C_F{}\else $C_F$ \fi}}
\def    \ca     {{\ifmmode C_A {}\else $C_A$ \fi}}

\def    \tf     {{\ifmmode T_F {}\else $T_F$ \fi}}
\def    \nf     {{\ifmmode n_f {}\else $n_f$ \fi}}
\def    \nc     {{\ifmmode N_c {}\else $N_c$ \fi}}
\def    \da     {{\ifmmode D_A {}\else $D_A$ \fi}}
\def    \Bf     {{\ifmmode B_F {}\else $B_F$ \fi}}
\def    \df     {{\ifmmode D_F {}\else $D_F$ \fi}}
\def    \b0{\mbox{$b_0$}}
\def    \MSB     {\ifmmode {\overline{\rm MS}} \else $\overline{\rm MS}$  \fi}
\def    \muf            {\mbox{$\mu_{\rm F}$}}

\def    \mur            {{\mbox{$\mu_R$}}}

\def    \mul            {{\mu_\Lambda}}

\def    \as             {\mbox{$\alpha_s$}}

\def    \assq           {\mbox{$\alpha_s^2$}}
\def    \ascube         {\mbox{$\alpha_s^3$}}

\def    \aemsq          {\mbox{$\alpha_{\rm em}^2$}}

\def    \e          {\ifmmode \epsilon \else $\epsilon$ \fi}
\def    \eps        {\ifmmode \epsilon \else $\epsilon$ \fi}
\def    \epsbar     {\ifmmode \bar\epsilon \else $\bar\epsilon$ \fi}
\def    \epsir      {\ifmmode \epsilon_{\rm IR} \else $\epsilon_{\rm IR}$ \fi}
\def    \epsuv      {\ifmmode \epsilon_{\rm UV} \else $\epsilon_{\rm UV}$ \fi}

\def    \feps#1         {\mbox{$f_{\epsilon}( #1 )$ }}

\def    \gaga           {{\gamma \gamma }}

\def    \Cggg           {\frac{1}{2M} \frac{\Phi_{(2)}}{3!}\frac{N}{K}}
\def    \Cggg2          {\frac{1}{2M} \frac{\Phi_{(2)}}{2!}\frac{N}{K}}

\def    \chizgg         {\mbox{$\Gamma(\chicz \to \gamma\gamma)$}}
\def    \chitgg         {\mbox{$\Gamma(\chict \to \gamma\gamma)$}}

\def    \chizlh         {\mbox{$\Gamma(\chicz \to \lh )$}}
\def    \chiolh         {\mbox{$\Gamma(\chico \to \lh )$}}

\def    \chitlh         {\mbox{$\Gamma(\chict \to \lh )$}}
\def    \pgg#1          {P_{gg}(#1)}
\def    \cpgq#1         {{\cal{P}}_{gq}(#1)}
\def    \cpgg#1         {{\cal{P}}_{gg}(#1)}
\def    \cpqq#1         {{\cal{P}}_{qq}(#1)}
\def    \cpqgamma#1     {{\cal{P}}_{q\gamma}(#1)}

\def    \o        {\ifmmode {\cal{O}} \else ${\cal{O}}$ \fi}
\def    \q        {\ifmmode {\cal{Q}} \else ${\cal{Q}}$ \fi}
\def    \oo       {\ifmmode \overline{\cal{O}} \else $\overline{\cal{O}}$ \fi}
\def    \oneSzero       {\ifmmode {^1S_0} \else $^1S_0$ \fi}
\def    \threeSone      {\ifmmode {^3S_1} \else $^3S_1$ \fi}
\def    \onePone        {\ifmmode {^1P_1} \else $^1P_1$ \fi}
\def    \threePJ        {\ifmmode {^3P_J} \else $^3P_J$ \fi}
\def    \threePzero     {\ifmmode {^3P_0} \else $^3P_0$ \fi}
\def    \threePone      {\ifmmode {^3P_1} \else $^3P_1$ \fi}
\def    \threePtwo      {\ifmmode {^3P_2} \else $^3P_2$ \fi}
\def    \spectrsh       {\mbox{$^{2S+1}L_J^{[1,8]}$}}

\def    \etah           {\mbox{$^1S_0^{[8]}$}}
\def    \etas           {\mbox{$^1S_0^{[1]}$}}
\def    \psih           {\mbox{$^3S_1^{[8]}$}}
\def    \psis           {\mbox{$^3S_1^{[1]}$}}
\def    \hcs            {\mbox{$^1P_1^{[1]}$}}
\def    \hch            {\mbox{$^1P_1^{[8]}$}}

\def    \chijh          {\mbox{$^3P_J^{[8]}$}}

\def    \chijs          {\mbox{$^3P_J^{[1]}$}}

\def    \jpsi           {\mbox{$J\!/\!\psi$}}
\def    \psip           {\mbox{$\psi'$}}

\def    \chicj          {\mbox{$\chi_{cJ}$}}

\def    \chicz          {\mbox{$\chi_{c0}$}}
\def    \chico          {\mbox{$\chi_{c1}$}}
\def    \chict          {\mbox{$\chi_{c2}$}}
\def    \QQ             {Q \overline Q}

\def\oppsih  {\mbox{$\langle 0\vert {\cal O}_8^{\psi}(^3S_1)\vert 0 \rangle$}}

\def\ophtpjs {\mbox{$\langle 0\vert {\cal O}_1^{H}(^3P_J)\vert 0 \rangle$}}
\def\ophoszs {\mbox{$\langle 0\vert {\cal O}_1^{H}(^1S_0)\vert 0 \rangle$}}

\def    \sp#1#2         {#1#2}
\def    \eik#1          { \frac{#1 \epsilon_c}{#1 k} }
\def    \eikg#1         { \frac{#1 \epsilon}{#1 k} }
\def    \ds#1           { \hat #1  }
\def    \lh             {\mbox{$\mathrm LH $}}

\begin{document}

\title{Quarkonium Decays and Production in NRQCD\footnote{
Invited talk presented at 5th Workshop on QCD,
Villefranche-sur-Mer, France, 3-7 Jan 2000}}

\author{Fabio Maltoni}

\address{University of Illinois at Urbana-Champaign,\\
1110 West Green Street, Urbana,IL 61801, USA\\E-mail:
maltoni@uiuc.edu}

\maketitle\abstracts{
        Some examples of the most recent applications of the NRQCD
        factorization approach to quarkonium phenomenology
        are presented. In the first part of the talk the NLO
        calculations for $\chicj$ and $\Upsilon$ decays
        rates are compared to the data and the results critically
        analyzed. In the second part, I show how information on the
        non-perturbative matrix elements can be extracted from
        the hadronic fixed-target experiments and eventually
        used to test their universality.}

\section{Introduction}
A large number\footnote{It will suffice to mention that the
seminal paper by Bodwin, Braaten and Lepage~\cite{Bodwin:1995jh}
has by now almost reached five hundred citations!} of
achievements, developments, and still-open questions enrich the
research activity on Non-Relativistic QCD; it would certainly be
hopeless to cover even a reasonable portion of them here.
Fortunately, there are several reviews available~\cite{reviews}
that we can all enjoy and I urge the interested reader to refer
to them. In this talk, I will therefore concentrate on some
specific examples which will give us the opportunity to see the
factorization approach of NRQCD at work. In so doing, it will
hopefully be easier to appreciate its strengths and to learn how,
whenever possible, to avoid its weaknesses. The results presented
in the following sections are mainly taken from Ref.~\cite{PHD}
and can be considered as applications of the next-to-leading
order (NLO) calculations performed in Ref.~\cite{Petrelli:1998ge}.

\section{Inclusive Quarkonium Decays}

The interest for studying quarkonium decays is two-fold. On the
one side we are faced with a lot of experimental data on masses,
total widths and branching ratios. On the other, we have at hand
predictions for these quantities, in most of cases with a NLO
accuracy. For instance, the annihilation of quarkonium into final
states consisting of leptons, photons and light hadrons can
provide useful information on the strong coupling at a scale of
the order of the quarkonium mass $M$. In general we can write the
inclusive decay width as an expansion in the NRQCD operators \ba
\Gamma\(H \to X \)\!=\! \sum_\q\, \hat\Gamma(\q \to x) \langle H
| \oo(\q)| H \rangle, \label{eq:Decay} \ea
\\[-8pt] where $\hat\Gamma(\q \to x)$
describes the short-distance decay of a $ \QQ $ pair in the
color, spin and angular momentum state $\q=\spectrsh$, while
$\langle H | \oo(\q)| H \rangle$ are the non-perturbative matrix
elements (MEs). Roughly speaking, they represent the probability
for the heavy-quark pair in the quarkonium $H$ to have quantum
numbers $\q$ when the annihilation takes place. Even if not
calculable within perturbative QCD, the MEs have definite scaling
properties with respect to the relative velocity in the
center-of-mass frame of the two heavy quarks,
$v$.~\cite{Lepage:1992tx} Therefore the sum in
Eq.~(\ref{eq:Decay}) has to be considered as a simultaneous
expansion in $\as(M_H)$ and $v$, and can be truncated at any
given order of accuracy.

There are several sources of uncertainties in these studies. The
first one is related to non-perturbative MEs that, at least for
colour-octet states, are poorly known. In fact, the idea of using
universality to gain information on the MEs is almost unfeasible
in inclusive decays. The main reason is that while in production
one can find processes where the contribution from a given
channel is enhanced such that values of particular MEs may be
obtained (\eg the high-\pt  $\jpsi$ production for $\oppsih$),
this is not possible in fully inclusive decays. This is the main
motivation for studying also semi-inclusive decays, where in
principle this kind of technique works (\eg the radiative decays
of Bottomonium
\cite{Brodsky,Field,Kramer:1999bf,Maltoni:1999nh}). Another
source of uncertainty is the renormalization scale dependence of
the results. As is well known, theoretical predictions {\it at
fixed order}  depend on the arbitrary renormalization scale
$\mur$, so for the short-distance coefficients we write \ba
\hat\Gamma\(\q\to x \)\!=\! A_\q\, \as(\mur)^{p_\q} \left[1 \!+\!
B_\q(\mur) \as(\mur)\!+ \! O(\assq)\right]  \label{eq:NLODecay}
\ea where $p_\q$ identifies the leading order, $B_\q$ is the NLO
correction and $\q$ is summed in Eq.~(\ref{eq:Decay}). Unless
$p_\q=0$, the coefficient $B_\q$ depends both on the scale $\mur$
and on the renormalization scheme. The natural expectation is
that the more terms are added to the perturbative expansion, the
less dependence on the arbitrary scale remains, and the stability
of the result improves.

In the following sections, I will discuss two cases of interest:
the charmonium $P$-wave decays and the $\Upsilon$ hadronic decay.
In the first case, we will see that, thanks to the measurements
available and to the symmetries of NRQCD, we can give a
satisfactory description of the $\chi_{cJ}$ decays and even
formulate a prediction of the total width of the $h_c$ (the
$^1P_1$ charmonium state). In the second example, we will instead
realize that the same strategy applied previously does not work
for $S$-waves and the data alone are not enough to constrain all
the unknown parameters. Nonetheless, we will find that {\it if }
the NRQCD expansion in the relative velocity $v$ works for
bottomonium (as it should), {\it then} the color-octet effects in
$\Upsilon$ hadronic decays are of the same order as the
color-singlet ones and cannot be neglected.

\subsection{Charmonium $P$-wave inclusive decays}\label{subsec:Pdecays}

One of the first applications and successes of the factorization
approach of NRQCD has been the analysis of the decays of the
$P$-waves.~\cite{Bodwin:1992ye} For such states the colour-octet
component is at the same order in the $v$ expansion as the
colour-singlet one: \ba
|\chi_{cJ} \rangle &=& O(1) |\chijs \rangle + O(v) |\psih g\rangle\,, \\
|h_c\rangle &=& O(1) | \hcs \rangle + O(v) |\etah g\rangle \,, \ea
and both Fock components have to be considered in computing
physical quantities. The necessity of including all the components
at a given order in $v$ to obtain a finite result is a general
feature of NRQCD. This is due to the  fact that the
non-perturbative matrix elements characterized by quantum numbers
conserved by the NRQCD interaction at a given order in $v$ mix
under renormalization group evolution. To leading-order accuracy
in $v$ we can use the heavy-quark spin symmetry, which relates
the matrix elements: \ba \langle\oo_1\rangle m_c^4 \equiv \langle
\chi_{cJ} |{\oo}_1(\threePJ)| \chi_{cJ}\rangle &=&
\langle h_c |{\oo}_1(\onePone)| h_c\rangle + O(v^2)\,,\\
\langle \o_8\rangle m_c^2 \equiv \langle \chi_{cJ}
|{\o}_8(\threeSone)|\chi_{cJ}\rangle &=&\langle h_c
|{\o}_8(\oneSzero)|h_c \rangle+ O(v^2). \label{eq:SpinSymmetryChi}
\ea Moreover, using the vacuum-saturation approximation, which
holds true up to corrections of $O(v^4)$, we can relate the
non-perturbative matrix elements in the electromagnetic decay to
the hadronic ones. As a consequence, the hadronic and
electromagnetic decay widths of $\chi_{cJ}$, at leading order in
$v$ and at NLO in $\as$, can be read from expressions in Appendix
C in Ref.~\cite{Petrelli:1998ge}  and rewritten in a compact form
as: \ba
\! \!\! \!\Gamma(\chi_{cJ} \to \lh ) &=& A_J \pi \assq \left(1+
\frac{\as}{\pi} B_J \right) \langle {\oo}_1\rangle + \pi \assq
\left(1+ \frac{\as}{\pi} D\right)
\langle \o_8 \rangle\,,\nn\\[-5pt]
&&\hspace*{6.7cm}(J=0,2) \;\\
\! \!\! \! \Gamma(\chi_{c1} \to \lh ) &=& A_1  \ascube \langle
{\oo}_1 \rangle + \pi \assq \left(1+ \frac{\as}{\pi} D\right)
\langle {\o}_8\rangle \;,\\
\! \!\! \!\Gamma(\chi_{cJ} \to \gaga ) &=& A_J^{\gaga} \left(1+
\frac{\as}{\pi} B_J^{\gaga}  \right) \pi e_Q^4 \aemsq \langle
{\oo}_1 \rangle\,, \; \;\qquad \qquad \;(J=0,2)
\label{eq:3PJtogaga}\\
\Gamma(\chi_{c1} \to \gaga) &=& 0 \,, \\
\! \!\! \!\Gamma(h_c \to \lh ) &=& A^h  \ascube \langle {\oo}_1
\rangle + \pi \assq C^h \left(1+ \frac{\as}{\pi} D^h \right)
\langle {\o}_8 \rangle \, ,\label{eq:1P1} \ea where $A,B,C,D$ are
known numerical coefficients which in general also dependent on
$L=2 b_0 \log \frac{\mur}{2 m_c}$, and we used the short-hand
notation $\as=\alpha_s^{\overline{MS}}(\mur)$. We note that the
hadronic width of the $\chi_{c1}$ is mainly due to the
colour-octet component.
%
{\renewcommand{\arraystretch}{1.2}
\begin{table}[t]
\begin{center}
\begin{tabular}{l|c|l} \hline\hline
 Process & $\Gamma$ $({\rm MeV})$ & Reference
\\
\hline
$\chizgg $& $(4.0 \pm 2.8) \times 10^{-3}$ & CBALL \cite{xball} \\
$\chizgg $& $(1.7 \pm 0.8) \times 10^{-3}$ & CLEO  \cite{CLEO-gg} \\
\hline
$\chitgg $& $(0.321\pm  0.095) \times 10^{-3}$ & E760  \cite{E760-gg} \\
$\chitgg $& $(0.7  \pm  0.3)   \times 10^{-3}$ & CLEO  \cite{CLEO-gg} \\
$\chitgg $& $(3.4  \pm  1.9)   \times 10^{-3}$ & TPC2  \cite{TPC2-gg} \\
$\chitgg $& $(0.44 \pm 0.08)   \times 10^{-3}$ & E835  \cite{E835a} \\
$\chitgg $& $(1.02 \pm 0.44)   \times 10^{-3}$ & L3 \cite{L3} \\
$\chitgg $& $(1.76 \pm 0.62)   \times 10^{-3}$ & OPAL \cite{Ackerstaff:1998ec} \\
\hline
$\chizlh $& $13.5 \pm  5.4$  & CBALL \cite{xball} \\
$\chizlh $& $15.0 \pm  4.9$  & E835  \cite{E835a} \\
$\chizlh $& $14.3 \pm  3.6$  & BES   \cite{BES98} \\
\hline
$\chiolh $& $0.64 \pm  0.10$ & E760  \cite{E760-tot} \\
\hline
$\chitlh $& $1.71 \pm  0.21$ & E760  \cite{E760-tot} \\
\hline\hline
\end{tabular}
\ccaption{}{\label{tab:data-chij-decay} Most recent experimental
results on $\chicj$ decay widths. The errors have been obtained by
combining in quadrature the statistical and systematic errors
given in the quoted references. The widths to light hadrons have
been obtained from total widths by removing the contributions of
known radiative decays (${\rm Br}(\chi_{c1} \to \gamma J/\psi) =
27.3 \pm 1.6 \% \;,\; {\rm Br}(\chi_{c2} \to \gamma J/\psi) = 13.5
\pm 1.1 \% $).~\cite{Caso:1998tx}}
\end{center}
\end{table} }
%
The result of a global fit of the experimental data
(Table~\ref{tab:data-chij-decay}) to the theoretical expressions
for the $\chi_{cJ}$ yields the set of  free parameters $\as,
\langle \oo_1 \rangle, \langle \o_8 \rangle $, summarized in
Table~\ref{tab:fit-results-0}. We verified that the extracted
value of $\as$, evolved using the two-loop $\beta$-function, is
compatible with the LEP average value
($\as(M_Z)=0.119$).~\cite{Caso:1998tx} From
Table~\ref{tab:fit-results-0} we can read off our best value for
the non-perturbative singlet matrix element\footnote{For the sake
of comparison with other results, the singlet matrix element is
given in the original normalization of Ref.~\cite{Bodwin:1995jh}.
($ {\oo}_1= \frac{\o}{2 N_c}$)} ($m_c=1.5 \;\gev$): \ba \langle
\chicj |{\o}_1(\threePJ)| \chicj \rangle /m_c^2 &=& (3.2 \pm 0.4
)\cdot 10^{-2} \;\gev^3\,. \label{eq:BestValues} \ea The above
value may be compared with the one obtained from the
Buchm\"uller-Tye potential \cite{Eichten:1995ch} $ \langle \chicj
|{\o}_1(\threePJ)| \chicj \rangle /m_c^2 =4.8 \cdot 10^{-2} \;
\gev^3$ and can be useful in determining the cross sections for
$\chicj$ production. As shown in Table~\ref{tab:fit-results-0} the
agreement between theory and data improves (\ie the
$\chi^2$/d.o.f. is smaller) after including the NLO corrections
and, within the assumed approximations, the NRQCD approach
appears to work quite well even for a heavy-quark as `light' as
charm.
{\renewcommand{\arraystretch}{1.2}
\begin{table}[t]
\begin{center}
\begin{tabular}{l|l|l|l} \hline\hline
 $\mu$ & $m_c$  & $1.5 m_c$ & $2 m_c$  \\
\hline
$\as$  $~~~$           $(10\%)$ & $ 0.320  $ & $0.293$& $0.273 $\\
$\langle \oo_1 \rangle$ $(10\%)$ & $ 2.67   $ & $2.41 $& $ 2.30 $\\
$\langle \o_8 \rangle$ $(20\%)$ & $ 2.35   $ & $1.93 $& $ 1.84 $\\
\hline
$\frac{\chi^2}{d.o.f.}$ & $ 15.2/10 $ & $15.0/10$ & $16.1/10$ \\
\hline\hline
\end{tabular}
\ccaption{}{\label{tab:fit-results-0} Results of the fit to the
non-perturbative MEs and to $\as$ for different values of the
renormalization scale. Values in ${\rm MeV}$. The statistical
errors from the fit are indicated in parenthesis. }
\end{center}
\end{table} }
%

{\it Sic stantibus rebus}, we can push the analysis further and
use the information gathered from $\chicj$ decays to obtain a
prediction for the hadronic width of the elusive $^1P_1$
charmonium state, the $h_c$. In order to curb the errors, we used
the results to a second fit on the $\chicj$ decay widths with
$\as$ fixed to the LEP average value. Using
Eqs.~(\ref{eq:SpinSymmetryChi}) and Eq.~(\ref{eq:1P1}), we obtain
the prediction for the hadronic decay rate of the $h_c$: \be
\Gamma(h_c \to \lh ) = ( 0.72 \pm 0.32 )\; {\rm MeV}\,, \ee where
the errors from the fit, from the unknown-higher order
corrections in $v$, and from the renormalization-scale dependence
have been combined in quadrature. A prediction for the total
width of the $h_c$ can also be given if the radiative width is
included: \be \Gamma_{\rm TOT} =\Gamma(h_c \to \lh  )+
\Gamma(h_c  \to \gamma \eta_c)\,. \ee We can estimate the
radiative width by assuming spin symmetry and using
~\cite{EiGo77} \be \Gamma(\onePone \to \gamma\, \oneSzero) =
\left( \frac{E_\gamma^h}{E_\gamma^\chi}\right)^3 \Gamma(\threePJ
\to \gamma\, \threeSone)\,. \ee Plugging
numbers~\cite{Caso:1998tx} into the above expression one indeed
verifies an approximate independence of $J$ and finds
$\Gamma(h_c  \to \gamma \eta_c) \simeq .45  \; {\rm  MeV}$. The
result for the total decay rate \be \Gamma_{\rm TOT}(h_c)= (1.2
\pm 0.3 )\; {\rm  MeV} \ee is somewhat higher, but still
compatible within errors with the experimental upper limit given
in Ref.~\cite{E760}, \ie $\Gamma^{h_c}_{\rm TOT} < 1.1 \; {\rm
MeV} \;  ( 90\% {\rm CL} ) $. We look forward to the new data on
$P$-waves that will hopefully  be collected in the near future by
the E835 experiment at Fermilab.~\cite{E835b}

\subsection{$\Upsilon$ inclusive hadronic decay}
\label{subsec:Upsdecay}

In the case of  $S$-waves, the spin-singlet states $\eta$ and the
spin-triplet (vector) states $J/\psi$, $\psi'$ and the
$\Upsilon(nS)$, the structure of the Fock state at order $v^4$ is
\ba
\vert \eta\rangle &=& O(1)\vert \QQ[\etas]\rangle +  O(v)\vert
\QQ[\hch]\rangle \nn\\
&+& O(v^2)\vert \QQ[\psih]\rangle
+ O(v^2)\vert \QQ[^1S_0^{[1,8]}]\,\rangle\, + O(v^6) \,,\\[10pt]
\vert V\rangle &=& O(1)\vert \QQ[\psis]\rangle + \sum_{J=0}^2
O(v)\vert
\QQ[\chijh]\rangle \nn\\
&+& O(v^2)\vert \QQ[\etah]\rangle + O(v^2)\vert
\QQ[^3S_1^{[1,8]}]\,\rangle\, + O(v^6)\,. \ea As we see, in
contrast to the $P$-waves, colour-octet states contribute only at
order $v^4$, and, at least for bottomonium, where $v^4\simeq
0.01$, we might expect that these effects should be safely
neglected. This conclusion is certainly true for spin-singlet
states when the magnitude of the short-distance coefficients is
considered: the LO contributions for $\etas, \etah $ and $\hch$
are all at $\assq$. There is no enhancement of the short-distance
coefficient and so color-octet effects for $\eta_{c,b}$ can be
safely neglected. Different is the case of spin-triplet states:
in fact, the LO contribution to the hadronic decay of a vector
colour-singlet state is at order $\ascube$ while the octets decay
already at order $\assq$. We will then include these terms, check
their relevance and eventually verify whether constrains on
colour-octet matrix elements can be obtained.

The total decay width can be written as: \ba \Gamma(V \to \lh )
&=&
  c_1(\psis) \langle V |{\oo}_1(\threeSone)| V\rangle \;\;+
  d_1(\psis) \langle V |\overline{{\cal P}}_1(\threeSone)| V\rangle\;\;\nn \\
 &+& c_8(\psih) \langle V |{\o}_8(\threeSone)| V\rangle \;\; +
    c_8(\etah) \langle V |{\o}_8(\oneSzero)| V\rangle \;\; \nn \\
 &+& \sum_J c_8(\chijh) \langle V |{\o}_8(\threePJ)| V\rangle \;\;.
\label{eq:GammaHadronic} \ea

Using the symmetries of NRQCD one can reduce the number of
independent MEs from seven to four. Nevertheless, it is clear that
too many MEs are present at order $v^4$ to be able to extract
detailed information from the experimental decay rate. An
additional constraint can be obtained from the width into two
leptons, to which only the color-singlet state contributes:
 \ba
\Gamma(V \to l^+l^- ) &=&
  c_{ll}(\psis) \langle V |{\oo}_1(\threeSone)| V\rangle \;\;+
  d_{ll}(\psis) \langle V |\overline{{\cal P}}_1(\threeSone)|
  V\rangle\;,
\label{eq:GammaLeptonic} \ea where $c_{ll}$ is now known even at
NNLO.~\cite{BSS98,CM98} The conclusion is that using the available
data, we can at most obtain a constraint on a linear combination
of three color-octet matrix elements. The relativistic
corrections to the singlet have been calculated numerically
in~\cite{KeMu83} and analytically in~\cite{Schuler94,LabEtAl} and
we make use of their results. Using the equations of
motion~\cite{GK} to relate the operator $\oo$ to the
$\overline{{\cal P}}$, and using heavy-quark spin symmetry to
reduce to one the unknown MEs of the octet $P$-waves, we obtain:
\ba \Gamma^{LO}(V \to \lh ) &=& \ascube \frac{40}{81}\(\pi^2
-9\)\(1 - \frac{M_\Upsilon -2 m}{m} \frac{19 \pi^2 -132}{12 \pi^2
-108}\)
\langle {\oo}_1(\threeSone)\rangle \nn \\
&+& \frac{\nf \assq \pi}{3}\langle {\o}_8(\threeSone) \rangle +
\frac{5 \assq \pi}{6}\(\langle  {\o}_8(\oneSzero) \rangle +
 7  \langle {\o}_8(\threePzero) \rangle\)\,,
\label{eq:V-LO} \ea where  the short-hand notation $\langle
{\o}(\{S,P\}) \rangle  \equiv \langle V |{\o}(\{S,P\})| V\rangle
/m^{\{2,4\}}$ has been introduced. Some comments are in order:

\begin{itemize}
\item
As anticipated, the ratio of the singlet short-distance
coefficient to that of the octets is typically $c_1/c_8 \leq
\as/\pi$ and the enhancement is quite effective.
\item
The relativistic corrections are very large for charmonium $(\sim
100 \%)$ and relevant for bottomonium $(\sim 25\% )$.
\end{itemize}
At NLO, the corrections for the octets are listed in Appendix C
of Ref.~\cite{Petrelli:1998ge}. For the singlet, the
short-distance coefficient can be found in Ref.~\cite{MaLe81}. To
study the effects of the NLO corrections we present some results
for a specific choice of the adjustable parameters. We fix
$\Lambda^5_{QCD}=237\; {\rm MeV}$ and $\mul=\mur= f m_b$ with
$f=1/2,1,2$ (from top to bottom), $m_b=4.8 \;{\rm GeV}$ and
$M_\Upsilon=9.46 \;{\rm GeV}$. The result for the LO expression
in Eq.~(\ref{eq:V-LO}) is the following:
\renewcommand{\arraystretch}{1}
\ba \Gamma^{LO}(V \to \lh ) &=& \left\{
\begin{array}{c}
$11 $\\
$5.5$\\
$3.1$
\end{array}
\right\} \cdot  10^{-3} \langle {\oo}_1(\threeSone)\rangle+
\left\{
\begin{array}{c}
$0.32$\\
$0.20$\\
$0.14$
\end{array}
\right\}
\langle {\o}_8(\threeSone) \rangle \nn\\
&+& \left\{
\begin{array}{c}
$0.20$\\
$0.13$\\
$0.087$
\end{array}
\right\} \left(\langle  {\o}_8(\oneSzero) \rangle+ 7 \langle
{\o}_8(\threePzero) \rangle \right), \ea
while for the NLO expression we get:
\ba \Gamma^{NLO}(V \to \lh ) &=& \left\{
\begin{array}{c}
$2.6$\\$4.0$\\$3.7$
\end{array}
\right\} \cdot  10^{-3} \langle {\oo}_1(\threeSone)\rangle+
\left\{
\begin{array}{c}
$0.27$\\
$0.26$\\
$0.22$
\end{array}
\right\}
\langle {\o}_8(\threeSone) \rangle \nn\\
&+& \left\{
\begin{array}{c}
$0.22$\\
$0.19$\\
$0.15$
\end{array}
\right\} \left(\langle  {\o}_8(\oneSzero) \rangle+ \left\{
\begin{array}{c}
$7.5$\\
$6.6$\\
$6.5$
\end{array}
\right\}\langle  {\o}_8(\threePzero) \rangle \right).
\label{eq:3s1-decay-nlo} \ea
As a result, the short-distance-coefficient's variation with
different choices of the renormalization scale is significantly
reduced at NLO. On the other hand, the improvement in our
theoretical predictions confirms that the typical ratio between
color-singlet and color-octet short-distance coefficient is
exactly of the same order of magnitude as the expected $O(v^4)$
suppression between the non-perturbative matrix elements. In other
words, unless unnatural cancellations between the various octet
contributions take place,\footnote{Remember that at NLO both
short- and long-distance coefficients become scheme dependent and
so, in general, non-perturbative matrix elements might not be
positive definite.} the color octet contributions are of the same
order as the color singlet one! This quite unexpected result
leads to the conclusion that the available extractions of
$\as(Q^2\simeq M_{\Upsilon})$ from the measured ratios of decay
widths~\cite{Caso:1998tx} ( such as  $ R_{\mu} = {\Gamma (\Upsilon
\to {\rm hadrons})}/ {\Gamma(\Upsilon \to \mu^+ \mu^-)} $ and
$R_{\gamma}={\Gamma( \Upsilon \to \gamma + X)}/{\Gamma
(\Upsilon   \to \mu^+ \mu^-)}$ ), {\it which neglect color-octet
effects}, should be regarded with great caution. (See, \eg the
discussion on color octet effects in the photon spectrum in the
decay $ \Upsilon \to \gamma + X$ in Ref.~\cite{Maltoni:1999nh}).
This is certainly one of the cases where it would be most useful
to have good estimates of the non-perturbative matrix elements
from lattice calculations.

\section{Charmonium Production and Universality}
\label{subsec:Hadroproduction}

In complete analogy with Eq.~(\ref{eq:Decay}), we can write the
cross section for producing a quarkonium state $H$ as a sum of
terms, each of which factors into a short-distance coefficient
and a long-distance matrix element:

\be d\sigma(H + X) = \sum_\q d\hat\sigma( \QQ [\q] +
x)\;\langle{\cal O}^H(\q)\rangle\, . \label{eq-fm} \ee

The matrix elements $\langle{\cal O}^H(\q)\rangle$ are related to
the probability for a $\QQ$ state with definite quantum numbers to
evolve into the quarkonium $H$. In contrast to the decays, the
large variety of processes in quarkonium production at both
fixed-target and collider experiments offers the concrete
possibility to thoroughly test the factorization approach and in
particular the universality of the non-perturbative MEs. In fact,
NRQCD teaches us that we can trade (up to $O(v^4)$ corrections)
the color-singlet matrix elements extracted from the decays for
the ones which enter in the production. Unfortunately this is not
feasible for color-octet matrix elements. Nonetheless, by
studying more exclusive processes, as the $p_T$ spectrum and/or
the polarization of the produced quarkonia at hadronic colliders,
it is possible to constraint single color-octet matrix elements
in the NRQCD sum of Eq.~(\ref{eq-fm}). Eventually we are just
left with one or two unknown parameters which can be constrained
quite well from the data.

In this section we briefly compare the results for
hadroproduction of charmonium at fixed-target experiments with
our predictions. Our analysis follows the lines of the one
presented in Ref.~\cite{Beneke:1996tk}: in particular we use the
same experimental data and just investigate the effects of the
inclusion of NLO corrections on the extraction of the MEs.

\begin{table}[t]
\addtolength{\arraycolsep}{0.2cm}
\renewcommand{\arraystretch}{1.5}
\begin{center}
\begin{tabular}{lccc}
\hline\hline
 ME &$\jpsi$ & $\psip$ & $\chi_{c0}$\\
\hline
$\langle {\oo}_1^H (^3S_1) \rangle  $     &   $1.16/6$& $0.76/6$ &  --\\
$\langle {\oo}_1^H (^3P_0) \rangle/m_c^2 \cdot 10^{2} $ & --  & -- & $4.0/6$\\
$\langle {\cal O}_8^H (^3S_1) \rangle \cdot 10^{2}$ & $1.06$ & $0.46$ &  $0.32$\\
\hline \hline
\end{tabular}
\end{center}
\ccaption{}{\label{tab:MEs} Independent MEs for charmonium
production used as input values in the fit. Other MEs can be
obtained by spin-symmetry relations. Values in ${\rm GeV}^3$. }
\end{table}
\begin{figure}[t]
\begin{minipage}[t]{0.45\textwidth}
\epsfig{figure=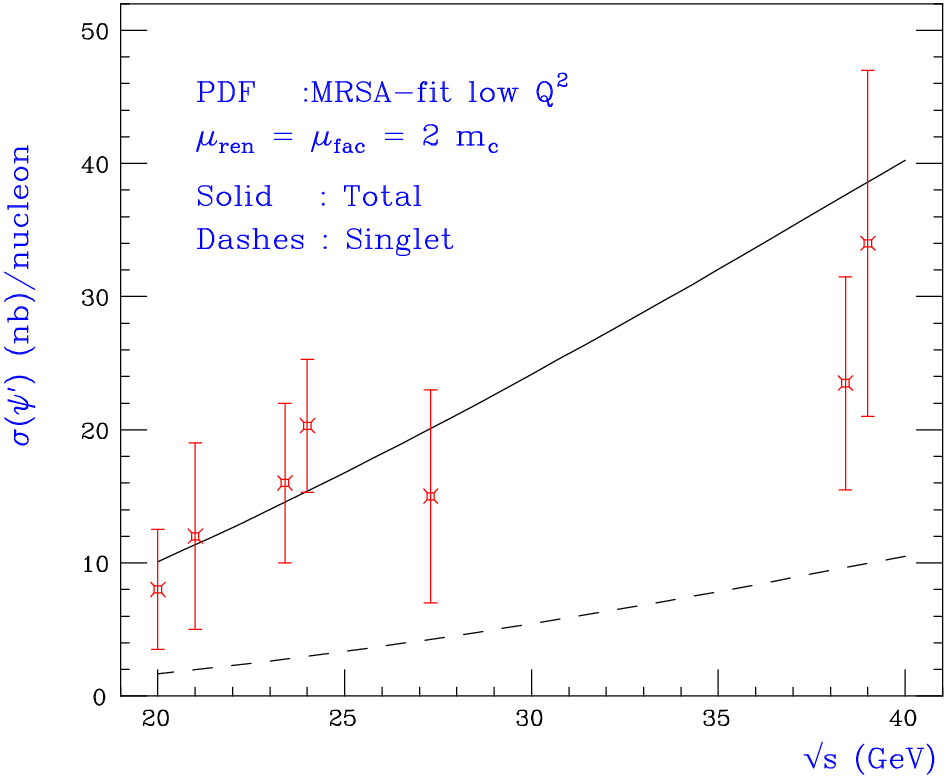,width=1.02\textwidth}
\end{minipage}
\qquad
\begin{minipage}[t]{0.45\textwidth}
\epsfig{figure=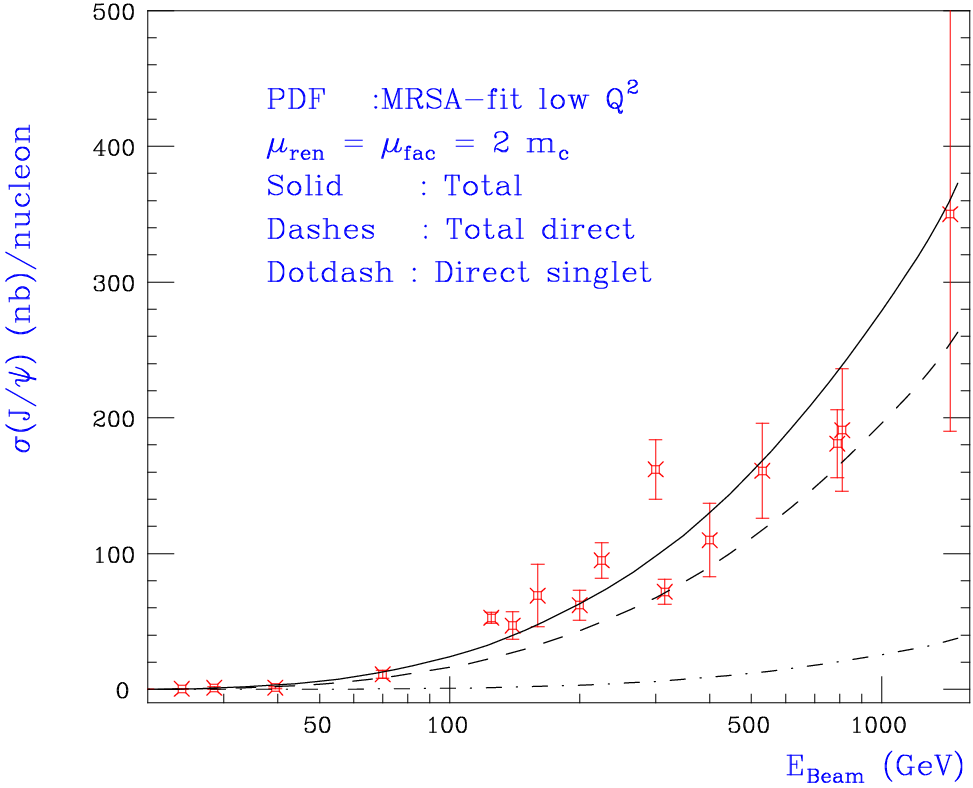,width=1.05\textwidth}
\end{minipage}
\ccaption{}{ \label{fig:fxt-psi} Left: Total $\psip$ (left) and
$\psi$ (right) production in proton-nucleon collisions ($x_F >0$
only). For the $\psi$, the total cross section includes the
radiative feed-down from $\chicj$ and $\psip$. The solid lines are
obtained with ${\cal M}^{\psip}_{6.4} = 2.6 \cdot 10^{-2}
\;\gev^3$ and ${\cal M}^{\psi}_{6.4} = 1.8 \cdot 10^{-2}\;\gev^3$
for $\psip$ and $\psi$ respectively.  }
\end{figure}
\begin{figure}[t]
\begin{minipage}[t]{0.45\textwidth}
\epsfig{figure=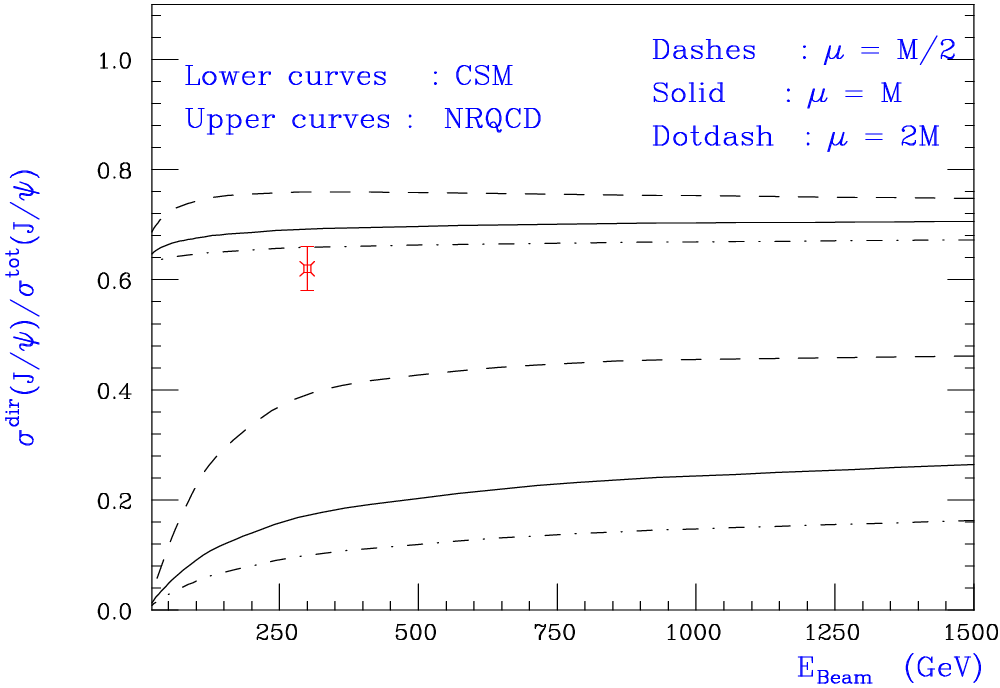,width=1.10\textwidth}
\end{minipage}
\qquad
\begin{minipage}[t]{0.45\textwidth}
\epsfig{figure=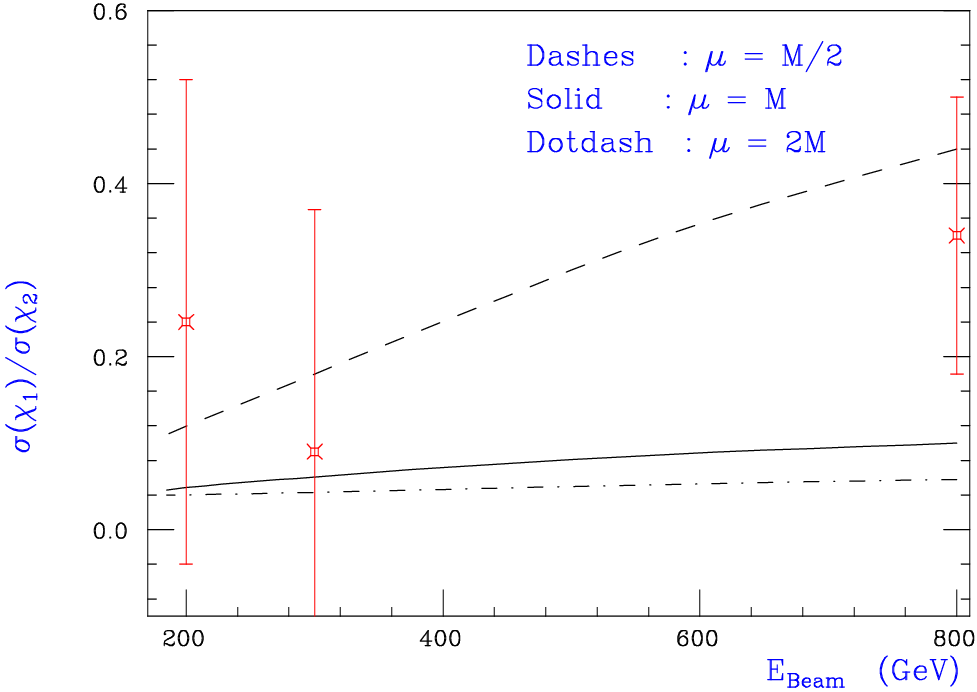,width=1.04\textwidth}
\end{minipage}
\ccaption{}{ \label{fig:ratios} Ratio of direct to total $\jpsi$
(left) and  $\chi_{c1}/\chi_{c2}$ (right) production in
proton-nucleon collisions as a function of the beam energy. The
experimental value at $300\; \gev$ for $\sigma^{\rm
dir}(\jpsi)/\sigma^{\rm tot}(\jpsi)$ is $0.62 \pm 0.04$. }
\end{figure}
In the following we have used the low-$Q^2$ MRSA set of parton
densities;~\cite{mrsaq2} $m_c=1.5\;\gev$ ; the renormalization,
factorization and NRQCD scales set equal to $\mur=\muf=\mul= 2
m_c$; the strong coupling, $\as$, tuned to the one used by the PDF
set, \ie $\Lambda_{\rm QCD}^5=152 \;\mev$; and the MEs collected
in Table~\ref{tab:MEs} as input values in the fits. Within the
above choices the only free parameter left is the linear
combination of MEs  \ba {\cal M}^{H}_k=\ophoszs+k\frac{
\ophtpjs}{m_c^2}, \ea with $k=7$ for the LO analysis and $k\simeq
6.4$ at NLO. The results of the fit are shown in
Fig.~\ref{fig:fxt-psi} and give:
\begin{equation}
{\cal M}^{\rm NLO}_{6.4}({}^1\!S_0^{[8]},{}^3\!P_J^{[8]})
=\left\{\begin{array}{c}
\,\,\,1.8\cdot 10^{-2}\,\mbox{GeV}^3\qquad (J/\psi) \\
0.26\cdot 10^{-2}\,\mbox{GeV}^3\qquad (\psi').
\end{array}
\right.
\end{equation}
to be compared with the LO results of~\cite{Beneke:1996tk}
\begin{equation}
{\cal M}^{\rm LO}_{7}({}^1\!S_0^{[8]},{}^3\!P_J^{[8]})
=\left\{\begin{array}{c}
\,\,\,3.0 \cdot 10^{-2}\,\mbox{GeV}^3\qquad (J/\psi) \\
0.52 \cdot 10^{-2}\,\mbox{GeV}^3\qquad (\psi').
\end{array}
\right.
\end{equation}
As expected, the inclusion of NLO corrections significantly
reduces the values needed for the color-octet MEs. Unfortunately,
the same sources of uncertainties present at LO, which are both
theoretical (the strong dependence of the results on the choice
of the heavy-quark mass and on the scale (which is only slightly
improved at these energies)) and experimental (nuclear
dependence, limited $x_F$ range, presence of elastically-produced
charmonia) make it difficult to associate reliable errors to the
fitted values.

Nevertheless, since most of these uncertainties should cancel in
the ratios of cross-sections, we have also compared \ba
R_{\chi}&=&\frac{\sigma(\chi_{c1})}{\sigma(\chi_{c2})}\qquad
R_{\psi}=\frac{\sigma^{\rm dir}(\jpsi)}{\sigma^{\rm tot}(\jpsi)}
\ea where \be \sigma^{\rm tot}(\jpsi)=\sigma^{\rm dir}(\jpsi)+
{\rm Br}(\psip \to \jpsi)\,\sigma(\psip) + \sum_{J} {\rm
Br}(\chi_{cJ} \to \jpsi) \,\sigma(\chicj)\,, \ee with the
available experimental data (see Ref.~\cite{Beneke:1996tk} for
details). These are shown in Fig.~\ref{fig:ratios}. While the
color-singlet approximation for $R_{\psi}$ is rather uncertain
and clearly disfavored by the data, both the LO and the NLO NRQCD
predictions show a much better agreement. Moreover, the NLO
result turns out to be quite stable under scale variations. With
a certain confidence, this result can be considered a remarkable
success of the NRQCD approach.

\begin{table}[t]
\addtolength{\arraycolsep}{0.09cm}
\renewcommand{\arraystretch}{1.1}
\begin{center}
\begin{tabular}{l|l|ccc|cc}
\hline\hline
 Process  &Ref.&$\langle {\cal O}_8^\psi (^3S_1) \rangle $& $k$ & ${\cal M}^{\psi}_k$& $\langle {\cal O}_8^{\psi '} (^3S_1) \rangle $& ${\cal M}^{\psi'}_k$\\
\hline
 Tevatron& \cite{CL96}                       & $0.66  $ & $3   $  & $6.6   $ & $0.46  $ & $1.8   $\\
         & \cite{Beneke:1997yw}              & $1.06  $ & $3.5 $  & $4.4   $ & $0.44  $ & $1.8   $\\
         & \cite{Cano-Coloma:1997rn}         & $0.3   $ & $ 3  $  & $1.4   $ & $0.14  $ & $0.33  $\\
         & \cite{Braaten:1999qk}             & $0.4   $ & $3.5 $  & $6.6   $ & $0.36  $ & $0.78  $\\
         & \cite{Nason:1999ta}               & $1.2   $ & $3.5 $  & $4.5   $ & $0.5   $ & $1.9   $\\
\hline
Fixed-target&\cite{Beneke:1996tk}                    & $(0.66)$ & $ 7  $  & $3.0   $ & $(0.46)$ & $0.52  $\\
hadropr.            &${^*}$                          & $(1.06)$ & $ 6.4$  & $1.9   $ & $(0.44)$ & $0.28  $\\
\hline
B-decay & \cite{Beneke:1999ks}               & $(1.06)$ & $3.1 $  & $1.5   $ & $(0.44)$ & $0.6   $\\
\hline \hline
\end{tabular}
\end{center}
\ccaption{}{\label{tab:Universality} Color octet matrix elements
for $\psi$ and $\psi'$ production as extracted from various
analysis ($ {}^*= $ this work).  Values in $10^{-2} \;{\rm
GeV}^3$. Numbers in parenthesis are taken as input values.}
\end{table}
The above analysis shows that, if on the one hand the NRQCD
approach gives a rather consistent description of the data, on
the other hand it is rather difficult to extract precise
information on the color-octet matrix elements from the
measurements. In this respect, it is interesting to compare some
of the extractions available in the literature with the one
above. The relevant color-octet matrix elements for the $\psi$ and
$\psi'$ are summarized in Table~\ref{tab:Universality}. Even if
the very basic assumptions of the cited works are often quite
different (PDF choice, inclusion of initial-state radiation and/or
$k_T$ effects, implementation of Altarelli-Parisi evolution,
etc.) and considering the theoretical uncertainties previously
mentioned, the overall picture is reasonably consistent.

\section{Conclusions}

I have presented some applications of the NRQCD formalism to the
phenomenology of inclusive quarkonium decays and production. In
$\Upsilon$ decays we have established the importance of
color-octet contributions to the total hadronic decay widths. We
have then argued that the present extractions of $\as$ from
$\Upsilon$ decays have to be regarded with great care, at least
until measurements of the relevant non-perturbative matrix
elements on the lattice are available. In the other two examples,
the $\chi_{cJ}$ decays and the charmonium production at
fixed-target experiments, we have been able to extract
information on the non-perturbative MEs directly from the data.
In the former case, this has allowed us to make a prediction for
the hadronic width of the $h_c$ charmonium state while in the
latter case, we checked the universality of the MEs against the
available results. The conclusion is that, considering the `light'
mass of the charm, the NRQCD approach works surprisingly well and
the extracted MEs are of the order of magnitude predicted by the
scaling rules. On the other hand, the uncertainties present both
at the theoretical and experimental level prevent us from
performing more precise tests of the most distinctive prediction
of NRQCD, \ie the universality of the non-perturbative matrix
elements. To overcome the above limitations, more exclusive
quantities are under study. The most exciting and promising one
is the expectation that high-$p_T$ quarkonia produced at hadronic
colliders should be transversely
polarized.~\cite{Cho-Wise,Beneke-Rothstein} Preliminary data from
CDF~\cite{CDF-pol} do not confirm this prediction and may indicate
that quarkonium physics is ready for new, unexpected surprises
(see, \eg Ref.~\cite{Lee-at-this-conference} and references
therein).

\section*{ Acknowledgments }
It is a pleasure to thank the organizers for an enjoyable
atmosphere and for financial support. I have benefited from
discussions and collaborations with Eric Braaten, Michael
Kr\"amer, Michelangelo Mangano and Andrea Petrelli. I am also
grateful to Diego Bettoni and Vaia Papadimitriou for their help
on experimental data. Finally, I thank  Scott Willenbrock for his
comments and suggestions on the manuscript. This work was
supported by the U.S.~Department of Energy under contract No. DOE
DE-FG02-91ER40677.

\end{document}